\documentclass[aps,prl,showpacs,twocolumn,groupedaddress]{revtex4}

\usepackage{amsmath}
\usepackage{amssymb}
\usepackage{graphicx}
  \graphicspath{{figures/}}
\usepackage{subfigure}
  
  \makeatletter\renewcommand{\@thesubfigure}{(\alph{subfigure})}\makeatother
\usepackage{overpic}

\begin{document}

\title{A spatially shifted beam approach to subwavelength focusing}

\author{Lo\"{i}c Markley}
\author{Alex M. H. Wong}
\author{Yan Wang}
\author{George V. Eleftheriades}
\email[]{gelefth@waves.utoronto.ca}
\affiliation{The Edward S. Rogers Sr. Department of Electrical and Computer Engineering, University of Toronto, 40 St. George Street, Toronto, Ontario, M5S 2E4, Canada.}

\date{\today}

\begin{abstract}

Although negative-refractive-index metamaterials have successfully achieved subwavelength focusing, image resolution is limited by the presence of losses.  In this Letter, a metal transmission screen with subwavelength spaced slots is proposed that focuses the near-field beyond the diffraction limit and furthermore, is easily scaled from microwave frequencies to the optical regime.  An analytical model based on the superposition of shifted beam patterns is developed that agrees very well with full-wave simulations and is corroborated by experimental results at microwave frequencies.

\end{abstract}


\maketitle

Subwavelength focusing of electromagnetic radiation has been the subject of a great deal of recent research in disciplines ranging from the microwave to the optical regimes.  When electric and magnetic fields are produced or scattered, the decay of evanescent field components away from their source or scatterer leads to a loss of spatial resolution below the order of one wavelength.  Many efforts have been made to overcome this effect, known as the diffraction limit, with various methods achieving a measure of success.  Early efforts based on the use of small apertures were very successful \cite{AshNicholls:1972}, leading to the field of near-field scanning optical microscopy (NSOM), but required extremely precise probe placements that increased the cost and complexity of the techniques.  In 2000, J. B. Pendry proposed an alternative strategy with a ``perfect lens'' that could focus electromagnetic waves to a point through the amplification of evanescent field components \cite{Pendry:2000}.  This was successfully demonstrated experimentally in planar form \cite{GrbicEleftheriades:2004} and in the quasistatic limit \cite{FangLeeSunZhang:2005,MesaFreireMarquesBaena:2005} but both cases suffered greatly from losses.  Other methods that have seen some recent developments are based on field synthesis through the plane-wave expansion of aperture fields and include direct field integration leading to near-field plates \cite{GrbicJiangMerlin:2008} and transmission screens inspired by holography \cite{EleftheriadesWong:2008}.  Here we show a different conceptual approach to synthesizing such transmission screens that is both simple and intuitive, incorporating the superposition of properly weighted dipole slot antenna field patterns.  The key enabling observation is that the beam patterns formed by displaced slot elements are spatially shifted in the near-field, allowing them to form a set of basis functions for a moment expansion synthesis of image-plane field distributions.  This technique leads directly to physically realizable implementations and experimental validation that is not sensitive to loss and can be readily scaled up to optical frequencies.

Field patterns with subwavelength features contain spatial Fourier components with transverse propagation vectors $k_{xz}$ larger than $k = \omega\sqrt{\epsilon\mu}$.  To satisfy Maxwell's equations, these components decay evanescently along $y$, and are therefore restricted to the near-field.  Negative-refractive-index slabs can be used to recover the near-field through evanescent field amplification but the image resolution tends to be very sensitive to losses \cite{Pendry:2000,GrbicEleftheriades:2004}.  Without such lenses, the image plane must remain close to the source since the field profile broadens quickly to the diffraction limit.  In this Letter, closely spaced dipole slots less than half a wavelength apart produce shifted beams that are weighted with out-of-phase currents to effectively increase the range in which the evanescent fields can be used to perform subwavelength focusing.

The electric field produced by an infinitesimally thin $z$-directed dipole slot antenna located at the origin is given in Eq.~\eqref{eqn:Ex} for a range of fields restricted to the $xy$-plane \cite{Balanis:2004}.  The slot is fed by a magnetic source current $I_m$, with the amplitude of the electric field related to the distance $\rho$ from the center of the dipole and the phase related to the distance $R$ from the tip of the dipole.  In accordance with the restrictions imposed by the experimental setup, only the $x$-component of the electric field is considered in this Letter, but any other component of the electric or magnetic field can be focused using the same principles outlined for $E_x$.

\begin{equation} \label{eqn:Ex}
    E_x(x,y,z) = \frac{I_m y}{j2\pi} \frac{e^{-jkR}}{\rho^2} = \frac{I_m y}{j2\pi} \frac{e^{-jk\sqrt{x^2 + y^2 + L^2/4}}}{x^2 + y^2}
\end{equation}

\begin{figure}
  \centering
  \includegraphics[clip]{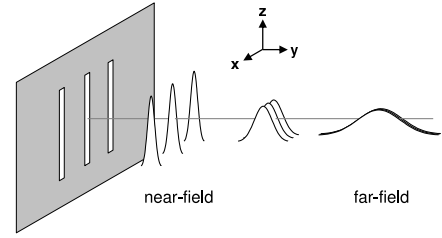}
  \caption{A qualitative illustration of how an array of dipole slots produces shifted beams in the near-field but identical magnitude distributions in the far-field.}
  \label{fig:shiftedbeams:qualitative}
\end{figure}

It is well known in optics and electromagnetics that the far-field distribution of a radiating system is simply the Fourier transform of the source field distribution \cite{YarivYeh:2007}.  This implies that if a slot were moved from the origin to a new position $x=d$ along the $x$-axis, the far-field pattern would experience only a phase shift, as shown in Eq.~\eqref{eqn:Exshiftedff}.  The separation of the far-field into the product of an element factor given by Eq.~\eqref{eqn:Ex} and an array factor is not valid in the near-field, however.  Indeed, the situation is distinctly different, giving us a new handle to manipulate the near-field for effects such as subwavelength focusing.  The near-field distribution of an antenna element translated along the $x$-axis is the spatially shifted beam described by Eq.~\eqref{eqn:Exshiftednf}.  The presence of strong evanescent waves produces fields that decay rapidly away from the slot element, leading to a localization of fields shifted away from the array origin.  This is critical in providing sufficient amplitude variation to design a subwavelength focused beam. As the distance from the source increases, the shifted beams spread until they can no longer be distinguished from one another strictly on the basis of amplitude and the property of spatially shifted fields is lost, as demonstrated in Fig. \ref{fig:shiftedbeams:qualitative}.

\begin{subequations}
    \begin{eqnarray}
        E_{x,shifted} & = & E_x(x,y,z)e^{-jkd\cos{\phi}} \quad \mbox{(far field)} \label{eqn:Exshiftedff} \\
        E_{x,shifted} & = & E_x(x-d,y,z) \qquad \quad \mbox{(near field)} \label{eqn:Exshiftednf}
    \end{eqnarray}
\end{subequations}

\begin{figure}
  \centering
  \includegraphics[clip]{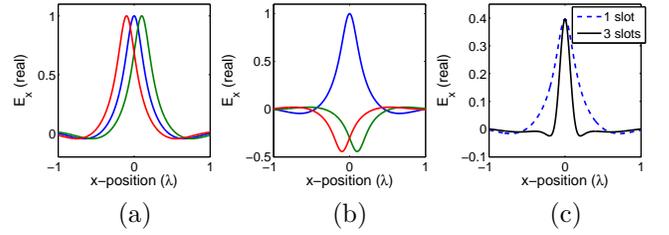}
  \caption{(color online) Subwavelength focusing from the superposition of shifted beams in the near-field.  The transverse electric field at $y=0.15\lambda$ is plotted at several conceptual stages.  (a) Shifted beams from slots at $x=-0.1\lambda$ (red), $x=0$ (blue), and $x=0.1\lambda$ (green).  (b) The shifted beams multiplied by their weighting factors. (c) The weighted shifted beams superposed to produce a subwavelength focused beam (solid) and the beam from a single slot (dashed).}
  \label{fig:shiftedbeams}
\end{figure}

For a source to focal plane separation of $y=0.15\lambda$, the individual dipole fields from a three-dipole array are plotted in Fig. \ref{fig:shiftedbeams}a.  By adjusting the complex weights on each of these element field profiles (Fig. \ref{fig:shiftedbeams}b), the two satellite slots can be designed to destructively interfere with the central profile and produce a null in the total field close to the peak of the main lobe (Fig. \ref{fig:shiftedbeams}c).  As the distance between the nulls decreases, the main lobe beam width becomes smaller and the beam is more focused.  A tradeoff is involved, however, since the element beam profiles are of finite width themselves, leading to increased side-lobe levels and higher main lobe attenuation relative to the constituent element fields.  Increasing the number of elements can serve to lower the side lobes but a limit is reached as to the minimum focused beam width for a given inter-element spacing $d$.

The analytical formulation for the beam produced by a single slot can be extended to a linear array with an arbitrary number of slot elements.  The total $x$-directed electric field for an array of $N$ elements is given by Eq.~\eqref{eqn:Extotal} where $w_n I_m$ is the $n^{th}$ weighted magnetic source current and $x=nd$ is the position along the $x$-axis of the $n^{th}$ slot.  In the vector representation, $\mathbf{w}$ is a $N$-element column vector containing the complex weights $w_n$ assigned to each slot and $\mathbf{E_x}$ is a $N$-element row vector containing the field from each element as a function of $x$.

\begin{eqnarray} \label{eqn:Extotal}
    E_{x,total} & = & \sum_{n=-(N-1)/2}^{(N-1)/2} w_n \frac{I_m y}{j2\pi} \frac{e^{-jk\sqrt{\left(x-nd\right)^2 + y^2 + L^2/4}}}{\left(x-nd\right)^2 + y^2} \nonumber \\
& = & \sum_{n=-(N-1)/2}^{(N-1)/2} w_n E_{x,n} = \mathbf{E_x} \cdot \mathbf{w}
\end{eqnarray}

The weights $w_n$ are chosen to scale the individual slot electric field components $E_{x,n}$ to produce the desired electric field profile at the focal plane.  In order to determine the optimal weights $\mathbf{w_{opt}}$ necessary to approximate a desired electric field profile $f(x)$, the individual slot field components can be regarded as basis functions for a method of moments field expansion.  By multiplying both sides of $f(x) \approx \mathbf{E_x} \cdot \mathbf{w_{opt}}$ by $E^*_{x,m}$ and integrating over $x$ we are left with a system of $N$ independent equations that can be solved for the $N$ weights $w_n$ by reducing the problem to that of inverting an $N \times N$ matrix.  Although these basis functions are linearly independent, they are not orthogonal so the off-diagonal terms of the matrix do not vanish and the analytic formulation becomes unwieldy.  Instead, $f(x)$ and $E_{x,n}(x)$ can be written as column vectors $\mathbf{b} = f[x]$ and $\mathbf{A} = E_{x,n}[x]$ by discretizing them over $x$, causing the method of moments to be reduced to the method of least squares \cite{FraleighBeauregard:1995} with an expression for $\mathbf{w}$ given in Eq.~\eqref{eqn:leastsquares}.

\begin{equation} \label{eqn:leastsquares}
        \mathbf{w_{opt}} = \left(\mathbf{A^H}\mathbf{A}\right)^{-1}\mathbf{A^H} \cdot \mathbf{b}
\end{equation}

The desired focusing effect is demonstrated using a Gaussian pulse with a full-width half-maximum (FWHM) beam width of $0.12\lambda$ as the target field profile for an array of three slots separated by $d=0.1\lambda$.  The optimal weights were calculated using Eq.~\eqref{eqn:leastsquares} to be $w_0 = 1$ and $w_{\pm 1} = 0.446 \angle 174^\circ$ for a focal plane at $y=0.15\lambda$ (Fig. \ref{fig:shiftedbeams}c).

\begin{figure}
  \centering
  \includegraphics[clip]{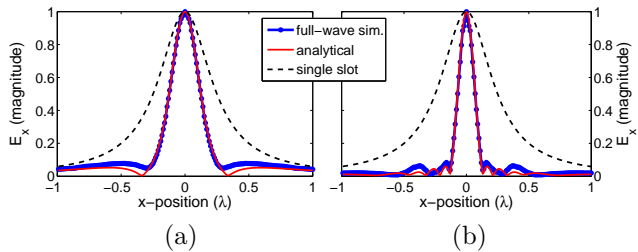}
  \caption{(color online) Simulated focused beams for slot arrays with 3 and 9 elements optimized for a focal plane at $y=0.25\lambda$.  The thin red line was calculated analytically from Eq.~\eqref{eqn:Extotal} while the thick blue line was extracted from a full-wave simulation.  The dotted line indicates the beam formed from a single slot.  (a) Three-element array with weights $w_0 = 1$ and $w_{\pm 1} = 0.497 \angle 172^\circ$.  (b) Nine-element array with weights $w_0 = 1$, $w_{\pm 1} = 0.764 \angle 179^\circ$, $w_{\pm 2} = 0.373 \angle {-}2^\circ$, $w_{\pm 3} = 0.136 \angle 175^\circ$, and $w_{\pm 4} = 0.032 \angle {-}6^\circ$.}
  \label{fig:minbeamwidth}
\end{figure}

In order to show the increased focusing power of an array with more elements, the focal plane was moved to $y=0.25\lambda$ where the beam width of a single slot is $0.5\lambda$.  The optimal weights were calculated for an $N=3$ and an $N=9$ array with $d=0.1\lambda$.  The structure was simulated using Ansoft's HFSS full-wave finite-element software with the half-wavelength dipole slots fed by ideal $x$-directed voltage sources.  The resulting field profiles matched very well with those calculated using Eq.~\eqref{eqn:Extotal}, converging to the analytical solution as the width of the simulated dipoles was reduced.  The three-element array was able to produce a beam width of $0.250\lambda$ (Fig. \ref{fig:minbeamwidth}a) while the nine-element array focused the beam further to a beam width of $0.144\lambda$ (Fig. \ref{fig:minbeamwidth}b).

In a numerical implementation, Eq.~\eqref{eqn:leastsquares} must be calculated using a truncated range of $x$ values, typically chosen such that the target field is very small at each endpoint.  When the weights are optimized over a small portion of the target function, however, such as by truncating the range of $x$ to those values falling within the target function's FWHM beam width, an arbitrarily narrow central beam can be formed for any $N \geq 3$ and any $d$.  However, this comes at the expense of extremely large side lobe amplitudes and high weight sensitivity, an observation reported in super-directive and super-oscillatory studies in the literature that can render ambitious applications of these procedures impractical \cite{FerreiraKempf:2006,DawoudAnderson:1978}.  If the weights are instead optimized over several wavelengths in $x$, the side lobes are reduced to reasonable levels but the narrow beam width is no longer maintained.  Increasing the number of elements reduces the side lobes further and allows a narrow beam width at more distant focal planes.

\begin{figure}
  \centering
  \includegraphics[clip]{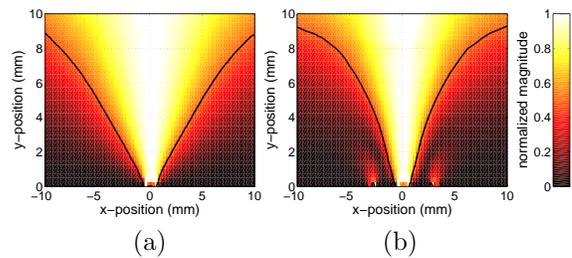}
  \caption{(color online) Simulated electric field magnitudes over the xy-plane for a single-slot (a) and a triple-slot (b) array under plane wave excitation at 10 GHz.  The fields are normalized along the $x$-direction with respect to the field value at $x=0$ to highlight the profile of the beam.  The thin black contour trace indicates the FWHM beam width.}
  \label{fig:xyplane}
\end{figure}

In order to realize this structure for a practical experiment, the ideal voltage sources at the centers of every slot were removed and plane wave incidence was considered.  As demonstrated through full-wave simulations in \cite{EleftheriadesWong:2008}, the weights were implemented by changing the dimensions of each slot, with a narrower slot producing a higher amplitude electric field and a wider slot producing a lower amplitude electric field.  Implementing the near $180^\circ$ phase shift between adjacent slots was more challenging and necessitated increasing and decreasing the lengths of adjacent slots to produce inductive and capacitive slots that could resonate together.  The final slot dimensions were optimized iteratively using full-wave simulations until the desired total field pattern was achieved.  Fig. \ref{fig:xyplane}a and Fig. \ref{fig:xyplane}b present a cross-section of the fields produced by the slots for a single-slot and a triple-slot arrangement, respectively.

\begin{figure}
    \centering
    \includegraphics[clip,scale=0.3]{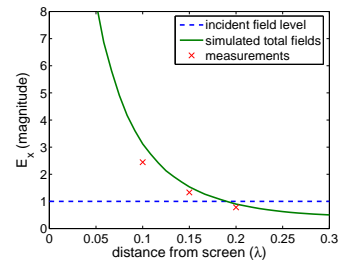}
    \caption{Full-wave simulation results for the magnitude of the electric field along the main axis of the focused beam as a function of distance from the screen.  The dashed line indicates the field strength of the incident electric field while the solid line indicates the field strength of the focused beam.  At the focal plane ($y=0.15\lambda$), the focused electric field strength is larger than the incident field by a factor of $1.54$. Measurement points from experiments conducted at $y=0.1\lambda$, $y=0.15\lambda$, and $y=0.2\lambda$ are included, showing a general agreement with simulations.  The magnitude of the measured field is normalized with respect to the background electric field (measured with the screen removed) and has a value of 1.33 at $y=0.15\lambda$.}
    \label{fig:EPAPS1}
\end{figure}

Although not visible in the normalized data of Fig. \ref{fig:xyplane}, despite the presence of rapidly decaying evanescent fields, the focused electric field magnitude at the focal plane is 1.5 times the magnitude of the incident field due to slot resonances (see Fig. \ref{fig:EPAPS1}).  The fields also show a similarity to those observed in negative-refractive-index slab focusing (see Fig. 3 and Fig. 4 from \cite{GrbicEleftheriades:2004}).  The presence of evanescent field components in any type of near-field focusing implies that rather than producing a three-dimensional spot, the fields are confined only in the transverse direction \cite{MesaFreireMarquesBaena:2005,GrbicJiangMerlin:2008}.  One major advantage of the focusing screens over metamaterial lenses, however, is the fact that Ohmic losses occur only at the screen and not throughout a region of wave propagation.  The structure is therefore insensitive to loss and although much of the incident power is reflected at the screen, this in no way affects the focusing behavior.  The simplicity of the slotted screen is also a distinct advantage over bulk metamaterials, allowing it to be easily scaled in frequency.

\begin{figure}
  \centering
  \includegraphics[clip]{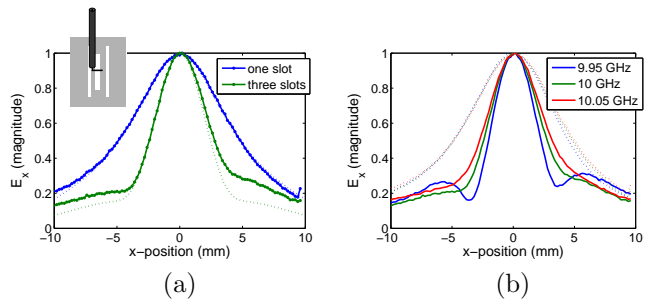}
  \caption{(color online) (a) Experimental measurements of the transverse electric field magnitude (solid lines) for single-slot and triple-slot arrays at 10 GHz compared with full-wave simulations (dotted).  The slight asymmetry in the measured field profile resulted from the asymmetrical coupling between the slots and the probe antenna (see inset).  (b) Measured field magnitudes at three frequencies for the single slot (dotted) and triple slot (solid) screens.}
  \label{fig:measurements}
\end{figure}

The dipole slots were created by cutting a three-dipole pattern in a 6 mil stainless steel sheet.  The slots were designed to operate at $10$ GHz with a central slot dimension of 13.2~mm $\times$ 1.2~mm and a satellite slot dimension of 17~mm $\times$ 0.6~mm spaced 3~mm away from the central slot.  An X-band horn antenna with a Rexolite biconvex lens was used to generate a collimated beam.  The slot apertures were placed at the Gaussian beam waist, forming a very good approximation to a plane-wave.  An Agilent E8364B network analyzer collected the field data over 8-12 GHz with one port connected to the horn antenna and the other port connected to a co-axial cable probe whose exposed inner conductor was bent ninety degrees to form a miniature dipole receiver antenna (see Fig. \ref{fig:measurements}a inset).  A custom built XYZ-stage from Newmark Systems positioned the probe over the screen with high precision to collect the near-field data on a plane 4.5 mm above the slotted sheet.

To verify the experimental setup and calibrate the probe, the fields produced by the horn and lens setup were measured with the screen removed and found to be collimated at the screen plane.  The transverse electric field was measured over the slots and the focused beam width was found to be 5.2 mm, which was within 6\% of the simulated value of $4.9$ mm.  The two satellite slots were then covered with 3M copper tape and the measurements were retaken, resulting in a single-slot beam width of 9.7 mm, within 1\% of the simulated value of 9.6 mm.  The measured beam width was reduced by a factor of 1.87 through the use of the two satellite slots (see Fig. \ref{fig:measurements}a).  In Fig. \ref{fig:measurements}b, two other frequencies close to the design frequency are plotted to show the frequency dependence of the focusing screen.

In summary, a shifted-beam approach to near-field focusing using slot antennas has been demonstrated in this Letter.  Using an array of dipole slots weighted out-of-phase with respect to one another, a set of shifted beams is formed that yields a subwavelength focused beam through superposition.  A moment-expansion method was developed to find the optimal weights given the number of elements, the element spacing, and the target field profile.  Out-of-phase weights were implemented for plane-wave excitation using resonating inductive and capacitive slots.  This led to experimental verification of the shifted-beam method for an $N=3$ array, with tighter focusing observed in simulations using larger arrays.

\begin{figure}
    \centering
    \includegraphics[clip,scale=0.3]{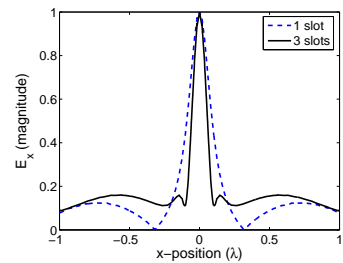}
    \caption{Simulation results from a full-wave simulation of the three-slot focusing screen adapted for optical frequencies ($\lambda = 830$~nm).  The transverse electric field is plotted at the focal plane $0.1\lambda$ away from the screen.  The metal screen was implemented using a lossy Drude model for silver with a finite thickness of 40~nm.  The slot lengths were shortened to 200~nm and 130~nm to compensate for the increased capacitance of slots at plasmonic frequencies.  The focused FWHM beam width was reduced to $63\%$ of the single slot beam width.}
    \label{fig:EPAPS2}
\end{figure}

The inherent scalability of the slot apertures provides the opportunity to implement the focusing screen at frequencies ranging from microwaves to Terahertz and all the way up to the optical domain for corresponding applications such as high-resolution imaging and sensing.  The dependence of the dipole slot behavior on plasma oscillations present in metals at optical frequencies can be readily accounted for by introducing an effective slot length that restores the scaled resonance frequency \cite{Novotny:2007}.  This was successfully demonstrated in simulations at $\lambda=830$~nm with slots cut into a 40~nm silver slab, as shown in Fig \ref{fig:EPAPS2}.  The increased subwavelength focusing range is very attractive for near-field imaging applications which are often constrained to working within an extremely small fraction of a wavelength away from the object to be imaged.  The structure demonstrated in this Letter can be used to resolve multiple objects in a variety of two-dimensional configurations with subwavelength precision by mechanically scanning the structure using orthogonal scanning patterns and detecting the back-scattered fields.  For complete imaging, the structure can be extended fully to two dimensions by adding a second set of satellite slots offset slightly along $x$ and shifted along $z$.

\end{document}